%% file: main.tex
\documentclass[letterpaper,twocolumn,10pt]{article}
\usepackage{usenix-2020-09}

\usepackage{tikz}
\usepackage{amsmath}  
\usepackage{pdfpages}
\usepackage{xurl}
\usepackage{float}
\usepackage{graphicx}
\usepackage{subcaption}
\usepackage{paralist}
\usepackage{booktabs}
\usepackage{graphicx}
\usepackage{subcaption}

\newcommand{\subhead}[1]{\vspace {0.5pt}\noindent{\textbf{#1.}}}

\def\Q{\vspace{-6pt}\list{}{\rightmargin 0.2cm\leftmargin 0.2cm}\item[]}

\newcommand{\new}[1]{\textcolor{black}{#1}}
\newcommand{\newer}[1]{\textcolor{black}{#1}}
\newcommand{\crs}[1]{\textcolor{black}{#1}}
\newenvironment{blue}{\par\color{black}}{\par}
\usepackage{soul}

\newenvironment{SUBENVccomment}[2]{\color{#1}[#2:]~}{\color{black}}

\definecolor{author1}{rgb}     {0.9,0.5,0.0}
\definecolor{author2}{rgb}     {0.6,0.0,0.8}
\definecolor{author3}{rgb}     {0.0,0.5,0.0}
\definecolor{author4}{rgb}     {0.9,0.2,0.2}

\begin{document}

\date{}

\title{\Large \bf ``My Privacy for their Security'': Employees' Privacy Perspectives and Expectations when using Enterprise Security Software\thanks{This paper has been accepted at Usenix Security Symposium 2023.}}



\author{
{\rm Jonah Stegman}\\
University of Guelph
\and
{\rm Patrick J. Trottier}\\
University of Guelph
\and
{\rm Caroline Hillier}\\
University of Guelph
\and
{\rm Hassan Khan}\\
University of Guelph
\and
{\rm Mohammad Mannan}\\
Concordia University
} 

\maketitle

\begin{abstract}
Employees are often \emph{required} to use Enterprise Security Software (``ESS'') on corporate and personal devices. ESS products collect users' activity data including users' location, applications used, and websites visited --- operating from employees' device to the cloud. To the best of our knowledge, the privacy implications of this data collection have yet to be explored. We conduct an online survey (n=258) and a semi-structured interview (n=22) with ESS users to understand their privacy perceptions, the challenges they face when using ESS, and the ways they try to overcome those challenges. We found that while many participants reported receiving no information about what data their ESS collected, those who received some information often underestimated what was collected.
Employees reported lack of communication about various data collection aspects including: the entities with access to the data and the scope of the data collected. We use the interviews to uncover several sources of misconceptions among the participants. Our findings show that while employees understand the need for data collection for security, the lack of communication and ambiguous data collection practices result in the erosion of employees' trust on the ESS and employers. We obtain suggestions from participants on how to mitigate these misconceptions and collect feedback on our design mockups of a privacy notice and privacy indicators for ESS. Our work will benefit researchers, employers, and ESS developers to protect users' privacy in the growing ESS market.\looseness=-1


\end{abstract}


\input{01-Introduction}
\input{02-Background}
\input{03-Design}

\input{04-OnlineSurveyDraft}
\input{05-Interview}
\input{06-Discussion}
\input{07-Limitations}

\input{08-Conclusion}

\bibliographystyle{plainurl}
\bibliography{ref}

\input{09-Appendix}

\end{document}

%% file: 01-Introduction.tex
\section{Introduction}
\label{sec:intro}

The proliferation of laptops and smartphones has profoundly changed how corporate 
data is stored, processed, and accessed. In the past, only a few employees were issued corporate laptops~\cite{ars:corporate_dev}. This practice has dramatically changed over the past decade as workers increasingly use corporate laptops for mobile work. On the other hand, a recent survey indicates that 85\% of US organizations are adopting a bring your own device (BYOD) policy for laptops and smartphones~\cite{bitglass:byod}. 
With the increased usage of corporate devices and mass adoption of BYOD, organizations have a strong interest to protect corporate data on these devices~\cite{downer_byod_2015}. 
To this end, various Enterprise Security Software (``ESS'') solutions are used to ensure endpoint security~\cite{chandel_endpoint_2019}. Endpoint Detection and Response (EDR) solutions are deployed to monitor employees using behaviour analytics~\cite{hassan2020tactical}. Secure Access Service Edge (SASE) is another approach that utilizes existing cloud infrastructure to manage devices containing corporate data at the network's edge~\cite{netskope_what_2021}. 
As expected, these controls collect and transmit data on users' activity for security analytics, \new{which may be subject to snooping by security analysts or data leakage.}

A recent survey of Malwarebytes readers (n=900) shows that 53\% of respondents reported sending or receiving personal email, 52\% read news, 38\% shopped online, and 25\% accessed their social media on their work device~\cite{malwarebytes:corporate_use}.  
User activity logging has serious privacy implications as all data, including personal, is collected~\cite{basinya_personal_2018}. 
Collection of user\footnote{We use the terms ``employee'' and (ESS) ``user'' interchangeably throughout this paper.} activity for employee monitoring for productivity purposes through ``bossware'' has resulted in employee outrage~\cite{employee_monitoring}.
Previous research focused on bossware, most notably the Electronic Frontier Foundation (EFF) analyzed and reported on bossware's invasive data collection practices~\cite{eff:bossware}. \new{These monitoring practices have skyrocketed in recent years and some tools now use monitoring data to assign security risk scores to employees~\cite{economist_hyper_surveilled, globe_surveillance}.}
Data collection for the purpose of enhancing enterprise security has largely remained uncontested and has not faced any legal challenges.
In the US, while email and Internet use are covered under the Electronic Communications Privacy Act (ECPA), there is an exception to it under the ``ordinary course of business'',  enabling employers to monitor email  or Internet use for a legitimate business purpose~\cite{friedman_workplace_2007}. 
Similarly, Articles 6 and 7 of the EU's General Data Protection Regulation (GDPR) arguably provide an exception for data collection for cybersecurity (``for the performance of a contract or legal obligation'' or ``for a task in the public interest'')~\cite{tolbert_will_2018,horak_gdpr_2019}. 
Several ESS websites mention ``helping clients'' with GDPR compliance, but lack any discussion on employees' privacy~\cite{truta_gdpr_2018,crowdstrike_gdpr_nodate,maloney_protecting_2017,okta_gdpr_nodate,palo_alto_networks_how_nodate}. 
While data collection for security is possibly justified, it should be coupled with well established practices of effective communication of privacy policies, informed consent, meaningful indicators and controls for users, and collection of only essential data.
However, to the best of our knowledge, privacy implications of employees' activity monitoring for enterprise security have not been explored. 


We conduct the first two-part study to quantitatively and qualitatively measure the established practices in this domain from employees' perspective.
The first part was a survey (n=258) to establish employees' understanding of the functionality of the ESS they used, and its data collection practices. We also used the survey to measure their comfort with different data collection practices, and whether ESS provided them with the necessary controls to manage their privacy. Next, we conducted semi-structured interviews (n=22) to build a deeper understanding of their privacy concerns, and how these concerns influenced their device usage behaviour for corporate and BYOD devices. We also asked  participants how their privacy could be improved in the context of ESS. 
Finally, we designed mockups for a privacy notice and several privacy indicators. The mockups were presented to participants and feedback was collected on the efficacy of interfaces and possible improvements for future revisions.

Our key findings include:

\begin{itemize}\itemsep0em
    \item The 258 survey respondents reported not receiving \emph{any} information from employers for 8\% of the 585 unique user interactions with ESS. 
    Follow-up interviews show that even when participants reported receiving some information on data collection, 59\% (13/22) of participants underestimated what was collected.  
    \item We identify participants' key concerns including their lack of knowledge of what was collected (i.e., which data sources) and when it was collected (always vs.\ when a potential anomaly was suspected). 
    \item 14\% (36/258) of the online survey respondents agreed that their ESS did not provide sufficient controls to manage their privacy. Similarly, 36\% (8/22) of interview participants emphasized the inadequacy of the existing indicators to communicate the presence of ESS.  
    \item We apply the interview findings to uncover several sources of misconceptions and report that  poor communication and ambiguous data collection practices lead to the erosion of trust between employees, ESS developers and employers.
    \item Participants' feedback on our mockups shows the effectiveness of some mockups and provides suggestions on how to improve others so that ESS developers and enterprises can work towards improving the state of privacy in the context of ESS.
\end{itemize}

%% file: 02-Background.tex
\section{Background and Related Work}


\subsection{ESS and Stakeholders' Perspectives}
\new{Various software or hardware solutions are used by enterprises to protect their assets. These solutions may have access to employee activities and behaviours for security monitoring purposes. We define ESS as security software that is installed on employees' personal or corporate devices (i.e., we exclude hardware firewalls). 
ESS provide different functionality such as Single Sign-On (SSO), personal firewalls, VPN client, or complete endpoint monitoring solutions that enforce the majority of organizations' security policies. 
\crs{Different ESS may provide different levels of controls to its users, which may depend on the type of ESS too. For instance, Netskope v94.1 provides users with the option to turn it on or off and view blocked events. Cisco AnyConnect v4.10 allows users to turn it on or off, disable captive portal detection, and block connections.} 
We briefly discuss  Endpoint Detection and Response (EDR) and Secure Access Secure Edge (SASE), as these two solution types capture and transmit more user activities for security purposes as compared to VPN clients or SSO.} 

EDR collects data on the endpoints, mines for malicious activities, and alerts corporate IT staff of any unusual activity~\cite{chuvakin_named_2013}.  The collected information includes network connections, processes, and/or websites visited, among others. The collected data and/or the alerts generated are logged and shared with security analysts within the organization, or an outsourced Managed Security Services Provider (MSSP). Prominent EDR solutions include Crowdstrike Falcon~\cite{crowdstrike_endpoint_2022}, and  Malwarebytes EDR~\cite{malwarebytes_malwarebytes_2022}. 
In case of SASE,  the traffic from the endpoints is routed through a corporate network that can be hosted/managed by a third party~\cite{lerner_say_2019, netskope_what_2021}. Through the use of TLS interception, a deeper insight into users' activity is gained, and potentially malicious operations are blocked, or sent to security analysts for reviews. Prominent SASE solutions include Netskope~\cite{netskope_homepage_2021}, and Zscaler SASE~\cite{zscaler_sase_2022}. 

The first stakeholder for ESS is the organizations, who are motivated to use these software solutions to protect corporate data on the employees' devices.
The data collected by ESS needs to provide rich insights for incident resolution. For instance, the resolution of an abnormal login alert for an employee from a new location requires the data on the usual/normal login location. Furthermore, this data often needs to be presented to a human analyst for the final verdict.
Another source of concern for organizations is the insider threat. A 2018 survey showed that 90\% of the surveyed organizations felt vulnerable to insider threats~\cite{crowds_research_partners_insider_2018}. High profile incidents such as the theft of Google's trade secrets by an employee also provide a strong reason for organizations to be cautious of their employees' activities~\cite{employee_theft}. 
The user and entity behavioural analytics (UEBA) may reveal the origins of data breaches in addition to reporting illegal or malicious actions undertaken by employees~\cite{shashanka2016user}. 
Finally, organizations are increasingly outsourcing their security operations to reduce their operational expenditure or better manage security~\cite{soc_outsourcing, kokulu2019matched}, which further complicates this problem as UEBA data needs to leave the source organization's infrastructure. 










The other stakeholder for ESS is the employees, who are subject to threats to their personal data (privacy threats due to employee monitoring for productivity are discussed in \S~\ref{subsec:related-work}). 
Employees are often required to use ESS on corporate devices to access corporate resources. 
They may be using corporate devices 
without going over the terms during installation/configuration, which has been reported by researchers for other software~\cite{schaub2015design, kotut2022tl, wilson2016crowdsourcing}. 
Employees are motivated to use BYOD to conveniently access corporate email or data (without requiring an additional corporate device to carry). 
Since employees are likely to use both corporate and BYOD devices for business and personal use~\cite{malwarebytes:corporate_use}, it is highly desirable from the employees' perspective to follow best practices in terms of communicating what data is collected, how it is stored, who has access to it, and the indicators and controls that show the boundaries of activities in different contexts. 

\subsection{Related Work}
\label{subsec:related-work}
\subhead{Bossware}
Productivity measuring software solutions (``bossware'') generally capture user activities to measure employee productivity and have also been promoted for improving employee performance~\cite{blackman_how_2020}.
\new{Bossware is installed on devices primarily used by non-IT professionals, e.g., healthcare, call centers, or insurance companies (see~\cite{eff:bossware}).}
Employee perceptions and threats to privacy of bossware have been the focus of several studies~\cite{chang_exploring_2015, chory2016organizational, kalischko2021electronic, kensbock2021big}.
There are also recorded incidents of these tools resulting in negative experiences, most notably an employee being terminated 
for not downloading such tools
on their personal device~\cite{news__school_2021}.
There is a clear negative sentiment against these tools from employees~\cite{employee_monitoring}. 
Previous surveys have shown that about half of the employees are against bossware as they are unsure whether their personal information will remain private~\cite{chang_exploring_2015}. Consequently, employees have expressed hesitation to use corporate devices~\cite{chang_exploring_2015}. 

While a clear negative sentiment towards bossware exists, ESS products are different as employers' objectives for data collection are better justified. However, no existing work has measured the perception of ESS from employees' perspective.

\subhead{Privacy Policies and Indicators}
Much research has focused on collecting permissions from users in the context of smartphones~\cite{felt2011android, felt_android_2012, wijesekera2015android}, web/social media applications~\cite{chia2012app, robinson2014cognitive, johnson2012facebook, bonneau2010privacy}, and IoT~\cite{tian2017smartauth, fernandes2016flowfence}. However, these efforts are only tangentially related to our work. In case of ESS, employees are expected to agree to provide \emph{all} data that is required for security and for performing their work. Therefore, we focus on the more closely related aspect of the presentation of privacy policies.



Applications often present Terms of Service (TOS) agreements and privacy policies to users at the time of installation or configuration~\cite{bohme_trained_2010}. Websites generally make this information available to users during their visit. Several research efforts have validated that users often skip the lengthy documents and accept  the agreements without reading them~\cite{obar_biggest_2020, gluck2016short}.
Consequently, much research has focused on when and how to present this information to the user~\cite{gluck2016short, schaub2015design, harkous2018polisis}. 

Most of the findings on how to present information to users are applicable for ESS, which we also leveraged when designing possible privacy notices and indicators  for our study. However, unlike other applications (e.g., healthcare or entertainment) where data collection may be fully controllable by the user, data collection in ESS is on behalf of the employer.
This top-down push to ``use'' the ESS hinders the  employee to pick an alternate. With limited options for the employees, the questions around when this data should be collected and how much insight should be provided to the employee about the collection, storage, and access to the data is important. 

A privacy friendly system should use effective indicators to convey that the data collection is in progress. Such indicators have been used for  webcams~\cite{portnoff2015somebody}, browser security controls~\cite{schechter2007emperor}, and privacy indicators for Android and iOS~\cite{egelman2015thing, aosp:privacy_indicators}. Several existing ESS products have a corresponding taskbar icon that may indicate whether the software is active. However, this icon may be hidden, or the employee may fail to notice it when using the device for personal use; the employee may not even understand what the tool is doing when it is active. While researchers have uncovered this ``tucked away'' affect for browser security indicators~\cite{schechter2007emperor, reeder2018experience}, no existing work has explored it in the context of ESS.

%% file: 03-Design.tex
\section{Study Design}
\label{sec:design}
\crs{We  investigate only privacy aspects of ESS and ignore the user experience of ESS. Our research questions include:} 
\begin{enumerate}\itemsep0em
    \item Do employees understand the data collection practices around ESS? If so, through which channel (ESS, employer, or self exploration)? 
    \item What are users' privacy perception of ESS? How those perceptions influence the behaviour of employees during their daily device usage?
    \item How do privacy perceptions for ESS differ between corporate devices and BYOD?
    \item Does ESS provide the right controls and privacy indicators to users to manage their privacy? If not, what are the right controls and indicators in this context?
\end{enumerate}

There are several challenges to this investigation.
First, different ESS may offer different features, or use different terminology to explain the same features. \new{These differences are compounded by the fact that users may comprehend features and data requirements of such features differently.} 
Second, users may perform different types of activities on corporate devices and BYOD. The feedback collected from the participants needs to take the device type into account. 
Finally, users may be required to use multiple ESS solutions (on the same or different devices) as several solutions complement each other. When eliciting feedback from participants, it is important to focus on the right ESS as well as users' understanding of how different ESS solutions work in conjunction.\looseness=-1

We conducted a two-part study. 
The first part employed an online survey that asked respondents about their ESS usage, and their understanding of the privacy implications of data collection. This helped us identify the channels used to communicate privacy aspects and quantify privacy concerns of the users.
To further enrich our understanding, and to find ways to mitigate the identified concerns, we followed up with a semi-structured interview with a subset of the survey respondents. 
We prioritized those participants who agreed to be contacted for the interview, had different levels of concern regarding privacy in ESS. 
These participants were invited for an hour-long interview to explore the sources of their confusion (if any), how ESS influenced their behaviours when they were using a device with ESS, and early feedback on the desirable controls and indicators to manage their privacy. 

\crs{Note that we conducted the online survey twice. The first iteration of the survey had two shortcomings: it did not have attention checks, and two Likert scales were unbalanced (i.e., neutral option was not in the middle). We addressed these limitations in the second survey. The interview participants were chosen after the first survey and these participants were not subjected to the second survey as these participants were debriefed. Therefore, the results for the online survey (\S~\ref{subsec:survey_results}) do not contain the responses from the interview participants.}

We document the recruitment process, study procedure, and results separately for the online survey (\S~\ref{sec:survey}) and semi-structured interview (\S~\ref{sec:interview}). The survey and semi-structured interviews were piloted with three participants and their feedback was used to revise our study instruments.  \new{We received approval from our institutional IRB, which required informed consent, PII anonymization, and allowing participants to withdraw their data up to two weeks after the study.}


%% file: 04-OnlineSurveyDraft.tex
\section{Online Survey}
\label{sec:survey}
\begin{blue}
The primary objective of the survey is to understand how users perceive the data collection practices and corresponding privacy implications while using ESS. 
We presented the participants with a list of 32 ESS solutions, which were curated from a keyword web search of terms ``endpoint detection'' and ``enterprise security tools'' (see the list in Appendix~\ref{sfl:vendors}). This list was used to assist participants and if their ESS was not listed, they could provide it as a free text. 
 We asked respondents to choose a maximum of three tools that they had experience with, though respondents may have used more ESS tools. This limit of three allowed us to collect quality feedback without overwhelming the subjects.

\subsection{Recruitment and Procedure}
Unlike bossware, ESS solutions are primarily used by skilled professionals. Getting participants with a professional background and familiarity with ESS through public sources (e.g., Kijiji or Craigslist) is challenging. Therefore, in addition to advertising the survey on regional Subreddits and Kijiji, we also sent personalized messages 
to our primary and secondary connections on LinkedIn and Facebook. 
\crs{The second iteration of our survey (after fixing the limitations of the first survey)} was hosted on two separate URLs. One URL was posted in the Reddit posts and Kijiji ad, and the other was shared directly by email or message with our connections. This allowed us to distinguish the recruitment source. \crs{237 participants took both versions of the survey. } 
Our approach has obvious limitations of convenience sampling, discussed in \S~\ref{subsub:demographics}. 

The survey participant criteria necessitated that the respondents currently use a security software provided by their workplace.  Participants responded to the survey on Qualtrics XM (see Appendix~\ref{app:online_survey}). The survey collected data on 
the following categories: (1) demographics and background; (2) ESS tools used; (3) understanding of the data collection by ESS; and (4) perspective on the sufficiency of privacy controls provided by ESS. 
Furthermore, to ensure the quality of responses, we employed Qualtrics' anti-ballot stuffing detection, a CAPTCHA, and attention checks built into the survey. 
Participants were compensated \$5 for the completion of the survey.\looseness=-1  

\begin{table}[t]
\caption{Online survey respondents' demographics}
\label{tab:demographicsOnline}
\centering
\small\addtolength{\tabcolsep}{-3.5pt} \renewcommand{\arraystretch}{0.75}
\begin{tabular}{ccccccccc} \hline \addlinespace[1mm]
\multicolumn{9}{c}{{\bf n = 258}} \\ \hline \hline \addlinespace[1mm]
\multicolumn{9}{c}{{\bf Gender} } \\
\multicolumn{3}{c}{\emph{Man}}  & \multicolumn{3}{c}{\emph{Woman}}  & \multicolumn{3}{c}{\emph{Other}} \\
\multicolumn{3}{c}{149}    & \multicolumn{3}{c}{109}    &  \multicolumn{3}{c}{-}  \\ \hline  \addlinespace[1mm]
\multicolumn{9}{c}{{\bf Country} } \\ 
\multicolumn{3}{c}{\emph{Canada}}  & \multicolumn{3}{c}{\emph{US}}  & \multicolumn{3}{c}{\emph{Other}} \\
\multicolumn{3}{c}{184}    & \multicolumn{3}{c}{74}    &  \multicolumn{3}{c}{-}  \\ \hline \addlinespace[1mm]
\multicolumn{9}{c}{{\bf Age (in years)}} \\ 
\emph{18--25}               & \emph{26--30}               & \emph{31--35}               & \emph{36--40}               & \emph{41--45}               & \emph{46--50}               & \emph{50+} & \multicolumn{2}{c}{\emph{Undisclosed}}  \\
20                 & 52                 & 99                 & 56                 & 13                 & 16                  & 2   & \multicolumn{2}{c}{-}\\ \hline \addlinespace[1mm]
\multicolumn{9}{c}{{\bf Self reported proficiency in IT}} \\
\multicolumn{3}{c}{\emph{Basic}} & \multicolumn{3}{c}{\emph{Intermediate}} & \multicolumn{3}{c}{\emph{Advanced}} \\
\multicolumn{3}{c}{42} & \multicolumn{3}{c}{162} & \multicolumn{3}{c}{54}   \\  \hline \addlinespace[1mm]
\multicolumn{9}{c}{{\bf Education or work in IT}} \\
\multicolumn{3}{c}{\emph{No}} & \multicolumn{3}{c}{\emph{Yes}} & \multicolumn{3}{c}{\emph{Undisclosed}} \\
\multicolumn{3}{c}{74} & \multicolumn{3}{c}{183} & \multicolumn{3}{c}{1}   \\  \hline \addlinespace[1mm]
\multicolumn{9}{c}{{\bf Highest education}} \\ 
\multicolumn{2}{c}{\emph{High school}}&
\multicolumn{2}{c}{\emph{Undergraduate}} & \multicolumn{2}{c}{\emph{Master's}} & \multicolumn{2}{c}{\emph{Doctoral}} & \multicolumn{1}{c}{\emph{Other}} \\
\multicolumn{2}{c}{23} & \multicolumn{2}{c}{127} & \multicolumn{2}{c}{84} & \multicolumn{2}{c}{22} & \multicolumn{1}{c}{2} \\ \hline 
\end{tabular}
\end{table}

\subsection{Results}
\label{subsec:survey_results}

\crs{For the first iteration of the online survey}, 428 participants were recruited through primary and secondary connections on LinkedIn who responded to the survey. 
In the second iteration of the survey (with added attention checks and balanced Likert scales),
in addition to advertisements on Subreddits and Kijiji, we requested the original 428 participants to retake the survey through emails (participants were remunerated for the second response). \crs{492 submissions were received for the second iteration.}
There are some limitations due to participants retaking the survey and we discuss those in \S~\ref{subsub:demographics}.


When reporting results, for test statistics, we use Pearson's Chi-Squared to compare categorical data, a Kruskal-Wallis one-way analysis of variance to compare Likert scale responses between respondents (e.g., their technology proficiency levels), and Wilcoxon Signed-Ranks test to compare Likert scale responses between questions. For multiple comparisons,  we use Bonferroni correction. 

 \begin{figure}[t]
    \centering
    \includegraphics[trim={15mm 25mm 35mm 25mm},clip, width=8cm] {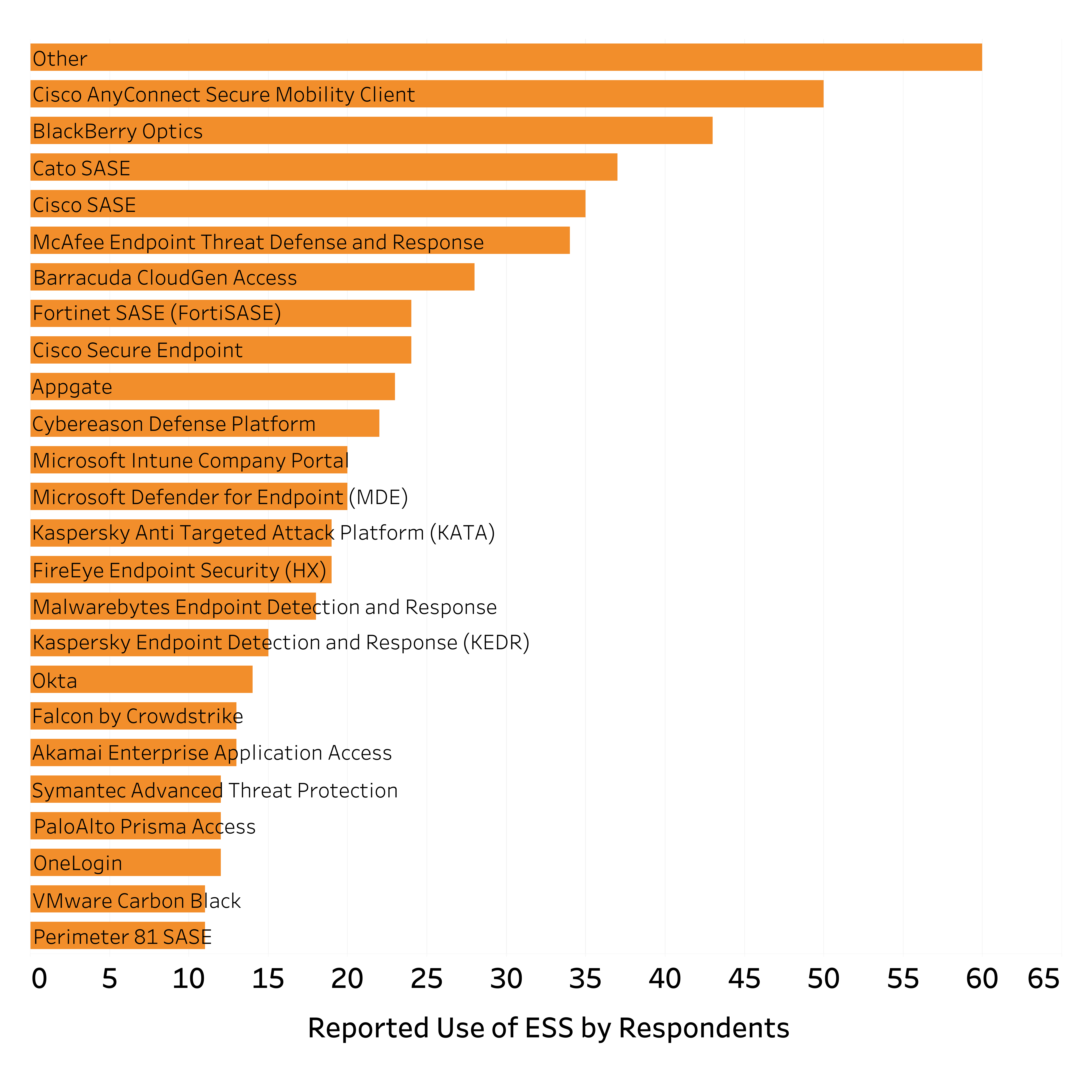}
    \caption{\crs{Number of times an ESS solution was reportedly used by survey respondents. ``Other'' includes ESS with ten or fewer selections and ESS beyond our listed solutions.}}
    \label{fig:tool_usage}
\end{figure}

\subsubsection{Demographics and Background}
\label{subsub:demographics}
We received 492 submissions of the survey. Results are reported for 258 responses that had no quality issues (attention checks were passed). 
21/258 respondents were from Kijiji and Reddit and 237/258 responses were collected through our LinkedIn and Facebook connections. 
\crs{The median time to complete the survey was 8 minutes and 40 seconds.} 
We asked respondents about their age, gender, country of residence, level of technology proficiency, industry type and size (asked for each ESS), their education, and background in technology. 

Table~\ref{tab:demographicsOnline}  summarizes the demographics of respondents. 42\% (109/258) of respondents were female, and 58\% (149/258) were male. Since the study was conducted in Canada, the majority of respondents are from Canada 71\% (184/258) and the rest are from the US. 
38\% (99/258) of the respondents are between the ages of 31--35 years, which  captures mid-career professionals. We also have representation from the other age groups: 8\% (20/258) between 18--25 years, 20\% (52/258) between 26--30 years, and 22\% (56/258) between 36--40 years of age. 
Both proficiency and education/work in IT show that over two-thirds of participants have experience with IT; 90\% (233/258) participants have at least an undergraduate degree.

\subhead{\crs{Convenience Sampling Limitation}}
We note representation from different demographics and different industries and organization sizes (reported in \S~\ref{subsec:ess_use}). However, due to our recruitment approach, we observe a large proportion of highly educated and IT focused respondents. 
This higher representation may also be expected due to the nature of ESS, which are primarily designed to protect technology resources of the organization.
Having more educated participants may help us identify privacy issues that may not be noticed by less educated or less IT-focused users. On the other hand, the latter group may have misconceptions around privacy in ESS (as evident by results reported in \S~\ref{subsec:privacy_perceptions}). 
Our results offer a likely upper bound on technical understanding and awareness of how ESS work and our work needs to be complemented with future studies that explore other population groups.

\subhead{\crs{Survey Retake Limitation}} 237/258 participants retook the survey because of an undiscovered flaw in our survey. We note that before retaking the survey, participants may have learned more about the topic. While this does not change our results on how the ESS or the employer communicated about the privacy aspects of the ESS, respondents may have learned about ESS on their own. We asked participants if they investigated the topic after the first survey attempt. 53\% (125/237) respondents reported that they investigated the topic on their own after the first survey attempt. It is likely that these participants established a deeper understanding of their ESS or this topic. 
\begin{table}[t]
\caption{Survey responses showing distribution of three information sources on how they communicated regarding what ESS does, features it provides, and data sources it uses. User-ESS interactions (n=585) in parenthesis.} 
\label{tab:channel_source}
\centering
\small\addtolength{\tabcolsep}{-1.0pt} \renewcommand{\arraystretch}{0.75}
\begin{tabular}{lrrr}
\hline \addlinespace[1mm] 
\multicolumn{4}{c}{\textbf{What it does?}} \\ 
                          & \multicolumn{1}{c}{\textit{ESS}} & \multicolumn{1}{c}{\textit{Employer}} & \multicolumn{1}{c}{\textit{Self}} \\
\textit{No communication} & 26\% (155)                               & 27\% (160)                                     & 28\% (167)                                 \\
\textit{Yes, somewhat}    & 48\% (280)                              & 40\% (237)                                   & 44\% (259)                                \\
\textit{Yes, clearly}     & 17\% (97)                              & 20\% (116)                                   & 17\% (102)                               \\
\textit{Don't remember}      & 7\% (40)                                & 11\% (64)                                     & 8\% (48) \\ 
\textit{Undisclosed}      & 2\% (13)                                & 1\% (8)                                     & 1\% (9) \\ \hline \addlinespace[1mm]
\multicolumn{4}{c}{\textbf{Features it provides}} \\ 
                          & \multicolumn{1}{c}{\textit{ESS}} & \multicolumn{1}{c}{\textit{Employer}} & \multicolumn{1}{c}{\textit{Self}} \\
\textit{No communication} & 24\% (139)                               & 25\% (147)                                    & 26\% (155)                               \\
\textit{Yes, somewhat}    & 33\% (195)                              & 38\% (223)                                    & 37\% (216)                               \\
\textit{Yes, clearly}     & 36\% (211)                              & 28\% (167)                                   & 28\% (165)                               \\
\textit{Don't remember}      & 5\% (29)                                & 6\% (37)                                     & 7\% (40) \\
\textit{Undisclosed}      & 2\% (11)                                & 2\% (11)                                     & 1\% (9) \\ \hline \addlinespace[1mm]
\multicolumn{4}{c}{\textbf{Data sources it uses}} \\ 
                          & \multicolumn{1}{c}{\textit{ESS}} & \multicolumn{1}{c}{\textit{Employer}} & \multicolumn{1}{c}{\textit{Self}} \\
\textit{No communication} & 27\% (158)                               & 30\% (175)                                    & 27\% (160)                                \\
\textit{Yes, somewhat}    & 34\% (199)                              & 37\% (219)                                   & 38\% (222)                               \\
\textit{Yes, clearly}     & 19\% (109)                              & 20\% (116)                                   & 18\% (104)                               \\
\textit{Don't remember}      & 18\% (108)                                & 10\% (60)                                     & 15\% (89) \\
\textit{Undisclosed}      & 2\% (11)                                & 3\% (15)                                     & 2\% (10) \\ \hline
\end{tabular}
\end{table}

\subsubsection{Reported ESS Use}
\label{subsec:ess_use}

The 258 respondents reported 585 unique user to ESS interactions (termed as \emph{user-ESS interaction}). 
30\% (78/258) respondents reported using one ESS, 12\% (31/258) reported using two ESS, and 58\% reported using three ESS. 
Note that we collect up to three user-ESS interactions per respondent. The distribution of usage across different ESS is summarized in Figure~\ref{fig:tool_usage}. All 32 ESS solutions listed were selected by one or more participants. 
Cisco AnyConnect Secure Mobility Client, BlackBerry Optics, Cato SASE, and Cisco SASE were the top reported ESS. Though ESS types fluctuate in use, we note a prominent use of SASE and EDR across participants (evident by ESS names).
For 85\% (499/585) of the user-ESS interactions, respondents reported currently using the ESS and another 8\% (45/585) reported that they used the ESS within past two years. 
Finally, 73\% (429/585) of the reported user-ESS interactions, respondents reported using the ESS for at least over a year.

In terms of the target device that ESS was installed on, 60\% (350/585) were reported on corporate devices, 25\% (149/585) on personal devices, and 14\% (80/585) on both corporate and personal devices. 1\% (6/585) interactions did not report this data. 59\% (153/258) respondents reported using ESS on at least one personal device. The usage distribution indicates that a significant proportion of ESS is used on  personal devices, which inevitably resides closely with users' personal data.

For each user-ESS interaction, we asked about the industry where the respondents were employed. 
The top four reported industries were Financial (23\%, 134/585), Healthcare (22\%, 129/585), Manufacturing (19\%, 114/585), and technology (15\%, 88/585). 
For the organization size, 15\% (90/585) of the reported user-ESS interactions were from organizations of size 1--10 employees, 26\% (150/585) from 11--50 employees, 33\% (191/585) from 51--100 employees, 19\% (112/585) from 101--500 employees, 4\% (21/585) from 501--1000 employees, and 2\% (13/585) from 1000+ employees.
These statistics show representation from different industries and  organization sizes for our survey's responses.


 \begin{figure}[t]
    \centering
    \includegraphics[trim={15mm 18mm 15mm 20mm},clip, width=7cm] {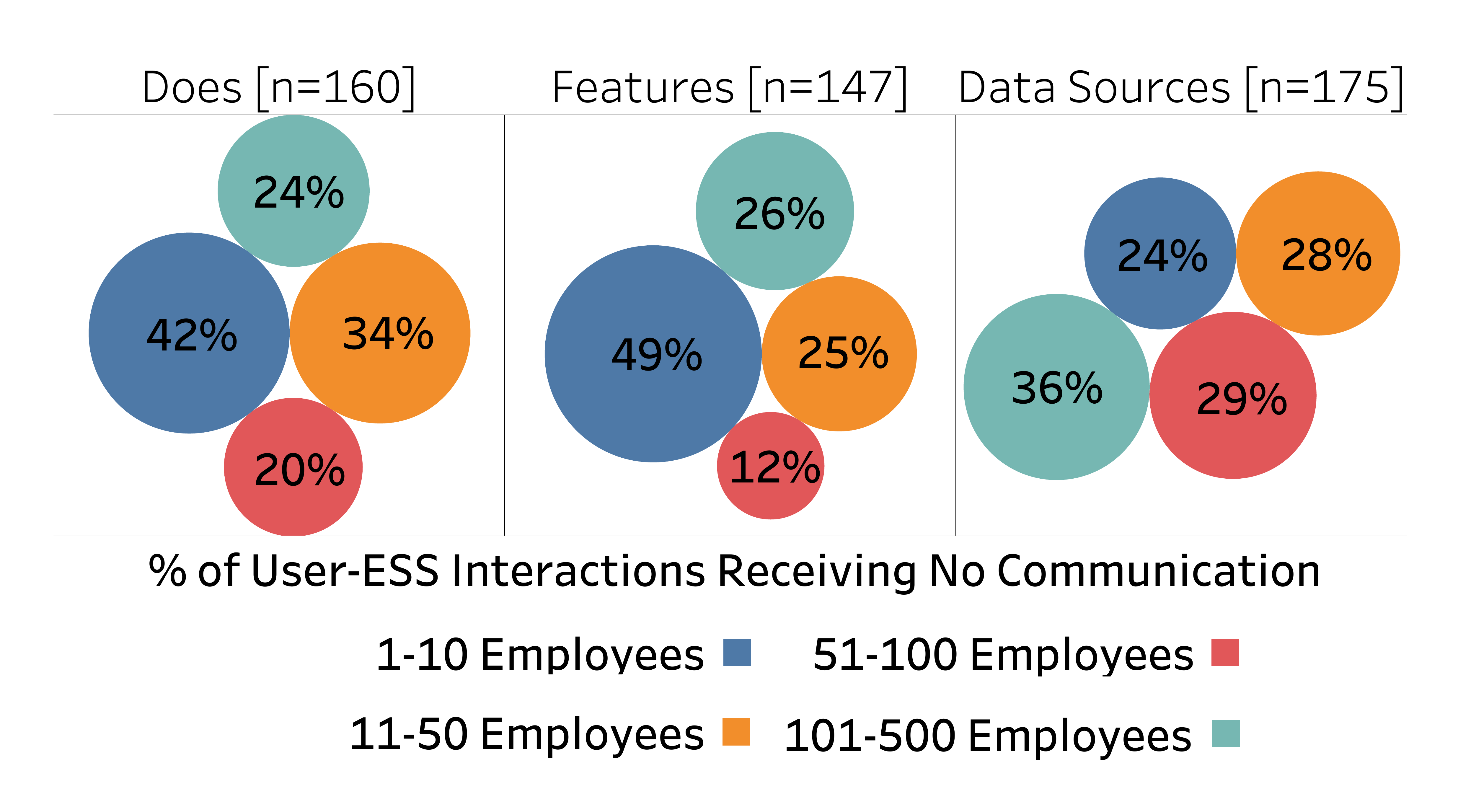}
    \caption{User-ESS interactions that received no communication 
    by the employer for different organization sizes (normalized by the organization size). Organization sizes 501--1000 and 1000+ are excluded due to limited samples.} 
    \label{fig:no_explain}
\end{figure}

 \begin{figure}[t]
    \centering
    \includegraphics[trim={17mm 55mm 20mm 20mm},clip, width=7cm] {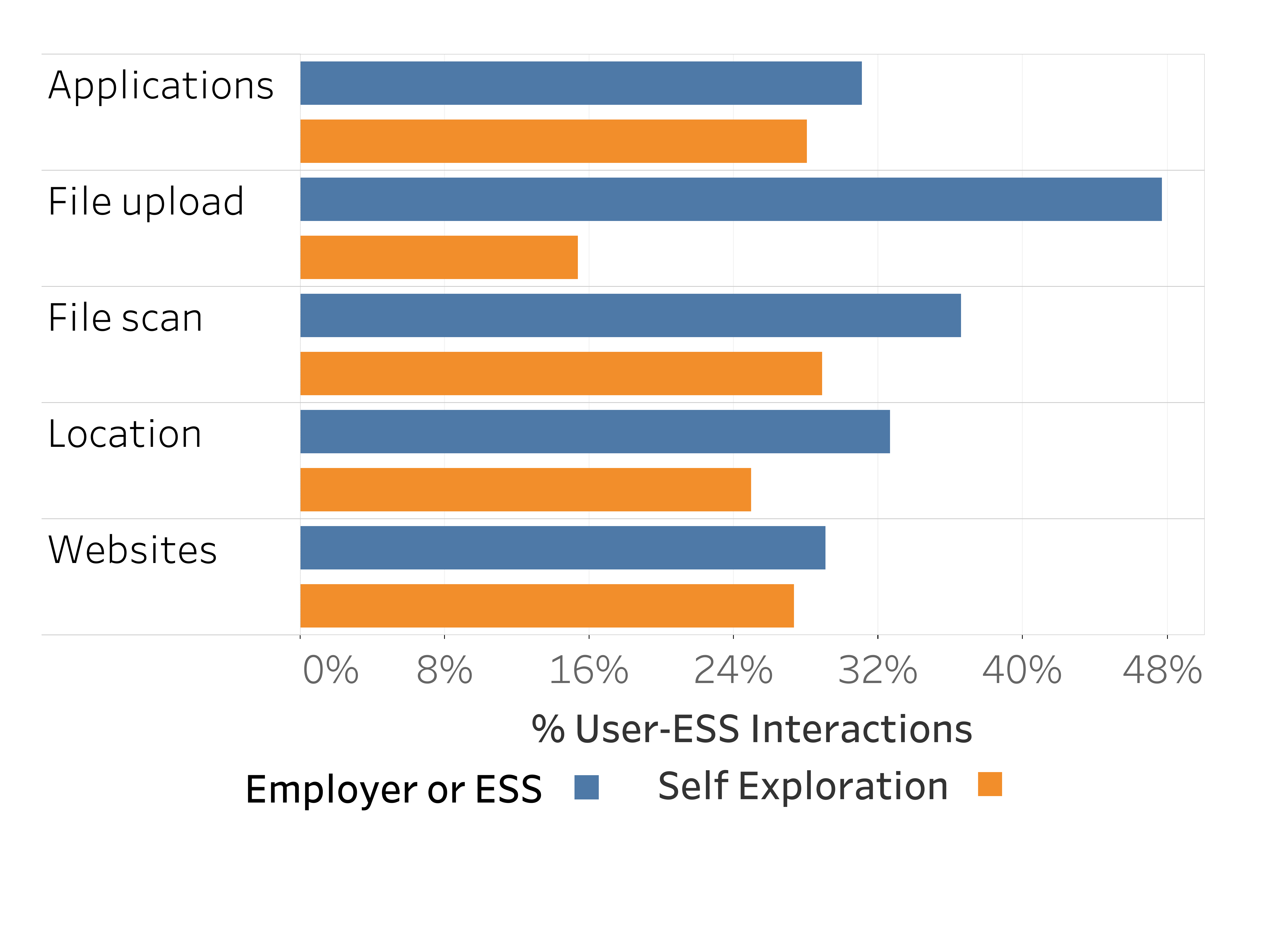}
    \caption{Percentage of user-ESS interactions where respondents reported learning about different data sources using self exploration or through ESS or employer.}
    \label{fig:data_source}
\end{figure}

\begin{figure*}[t]
   \centering
   \includegraphics[trim={0mm 43mm 5mm 28mm},clip, width=10cm] {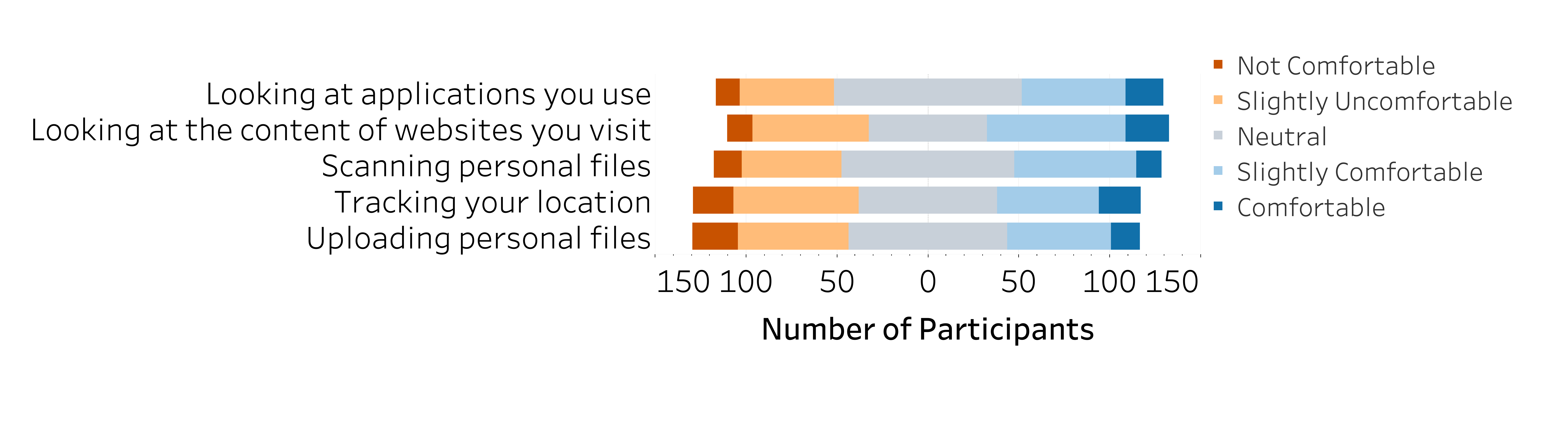}
   \caption{Reported comfort of survey respondents to different activities by ESS.}
   \label{fig:privacy_concerns}
\end{figure*}

\subsubsection{Users' Understanding of Collected Data}
\label{sec:user_understanding_collecti on}

We look at users' understanding of ESS data collection from two perspectives: where do they receive this information from which we define as ``information sources'' 
and what is conveyed through these sources. 
The three information sources that we considered are: ESS (e.g., through a privacy notice), employers (e.g., through training), and users' self-exploration (e.g., from online sources or colleagues). In terms of the information conveyed, key aspects include what the software does (i.e., purpose of software---e.g., ``lets me access corporate resources''), features it provides, and data sources that it uses (i.e., where it collects data from---e.g., ``collect data from websites visited''). 
Note that respondents' understanding of these aspects may be flawed, which we discuss further 
in \S~\ref{sec:limitation}.

Table~\ref{tab:channel_source} shows whether the three possible information sources (ESS, Employer, and Self), informed users about what the ESS tool does, features it provides, and data sources it uses for 585 reported user-ESS interactions.
In terms of ``what it does'', for at least 26\% (155/585) of the user-ESS interactions, respondents were not informed by one of the three sources. 
We note a similar pattern for ``provided features'' and ``data sources''--- at least 24\% (139/585) and 27\% (158/585) had no knowledge from one of the information sources.
For all three questions, we see that more respondents were not informed or do not remember than of those who had a clear understanding across all three information sources.
For 2\% (13/585) user-ESS interactions, twelve unique respondents received no information about the features that the ESS provides from any source.  The remaining user-ESS interactions were informed by one or more of the three sources. 
We note that on average, across the three questions, 39\% (228/585) respondents choose (``Yes, somewhat''), which indicates less certainty about the information that they received.
A Bonferroni-corrected Chi-squared test found no significant effect for all three questions about ESS (what ESS does, features it provides, and data sources it uses) and whether they received the information about the question (all $p$ > $0.01$). 

Figure~\ref{fig:no_explain} shows the distribution of user-ESS interactions where the respondents received no information from their employer for different organization sizes. The results are reported for the type of information conveyed (i.e., what ESS does, features it provides, and data sources that it uses) and are normalized for the organization size. 
Chi-squared tests found  significant effect for the organization sizes and whether respondents received no communication for ``What it does'' and ``features it provides'' ($p~<~0.01$ (Bonferroni-corrected)).
Post hoc comparisons show that respondents from organizations of size 1--10 and 11--50 employees are more likely to receive no information for the two questions than those of with 51--100 employees (both $p~<~0.008$). 
%
While certain factors may explain this difference---e.g., organization type (technology vs.\ manufacturing) or less maturity of organizations with fewer  employees---we did not collect data to find a plausible explanation. This is a possible avenue for a future work.\looseness=-1  

In Figure~\ref{fig:data_source}, we report percentage of ESS-user interactions where respondents reported learning about different data sources (e.g., location tracking) using self exploration vs.\ ESS or employer. 
A Chi-squared test found significant effect for the five data sources and whether they received the information by self exploration or employer/ESS  ($\chi^2(4) = 57.7, p~<~0.01$).
Bonferroni-corrected post hoc comparisons show significant differences between ``File upload'' and others (all $p~<~0.005$). Respondents were more likely to learn about ``File upload'' through their employer or ESS. 

\begin{figure}[t]
   \centering
   \includegraphics[trim={15mm 14mm 15mm 50mm},clip, width=6cm] {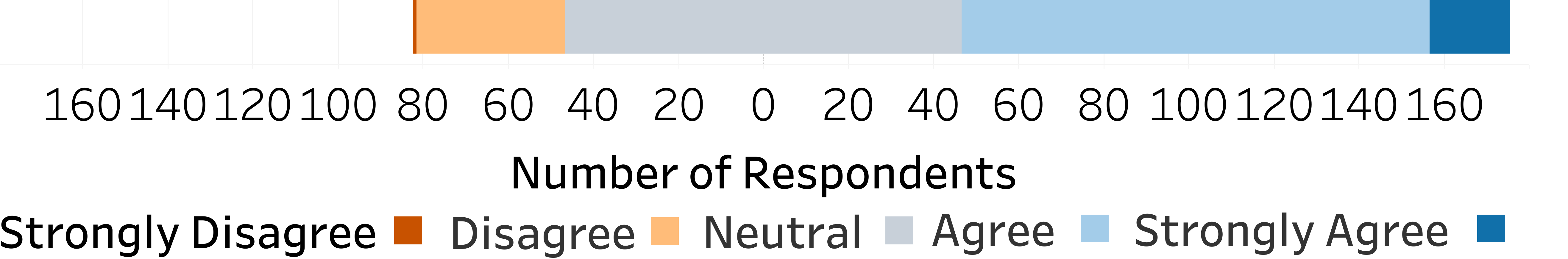}
   \caption{\crs{Survey response to: ``Do you feel that the enterprise security tools, such as the tools you chose previously, provide you with sufficient controls to manage your privacy preferences?''}}
   \label{fig:provides_privacy}
\end{figure}

\subsubsection{Privacy  Perceptions}
\label{subsec:privacy_perceptions}

To further understand participants perspectives around ESS privacy, first we asked their comfort for each of the five types of data collection activities by ESS on a 5-point Likert scale. These activities include: checking applications used, websites visited, scanning of personal files, location tracking, and uploading 
personal files of a user.
Figure~\ref{fig:privacy_concerns} 
shows the responses for each data collection activity. 
For respondents' comfort across five data collection activities, on average 7\% (18/258) reported being not comfortable, 23\% (60/258) slightly uncomfortable, 33\% (85/258) neutral, 24\% (63/258) slightly comfortable, and 8\% (20/258) comfortable, respectively.
The majority of the responses for all five activities are in the range from ``Neutral'' to ``Comfortable''. 
While all respondents were neutral or comfortable with one or more data collection activities, 27\% (70/258) were uncomfortable with at least one data collection activity.

We asked respondents if the ESS provides sufficient controls to manage their privacy preferences.
Figure~\ref{fig:provides_privacy} shows their responses on a 5-point Likert scale (``Strongly Disagree'' to ``Strongly Agree''). It shows that 50\% (129/258) strongly agree or agree, 36\% (93/258) are neutral, and 14\% (36/258) disagree or strongly disagree that ESS provides sufficient controls to manage their privacy preferences. \crs{The online survey's first iteration resulted in 54\% (231/428) strongly agree or agree, 37\% (158/428) neutral, and 9\% (39/428) disagree or strongly disagree.}
A Kruskal-Wallis test examined the effect of respondent technical proficiency on whether they feel that ESS provides sufficient controls to manage their privacy and found significant differences ($H(X) = 229.8, p~<~$0.05). 
The responses show that while the majority of respondents with basic technology proficiency were neutral 55\% (23/42), those with intermediate proficiency agreed 41\% (66/162), and those with advance proficiency also agreed 55\% (30/54) that ESS provide sufficient controls. \new{One possible reason is that the respondents with basic technology proficiency are unable to find these controls.} 

\begin{figure}[t]
   \centering
   \includegraphics[trim={10mm 25mm 10mm 33mm},clip, width=6cm] {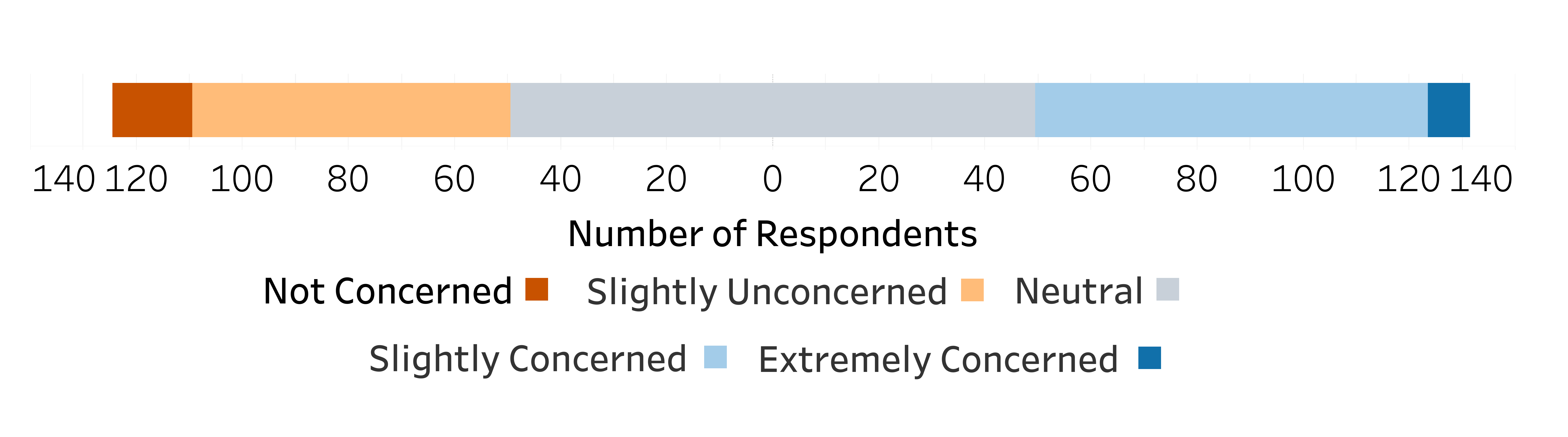}
   \caption{Survey response to: ``Are you concerned about your privacy when using enterprise security tools such as the tools you chose previously?''}
   \label{fig:concern_privacy}
\end{figure}

Finally, respondents were asked how concerned they are with respect to their privacy when using ESS tools. Figure~\ref{fig:concern_privacy} shows that 6\% (15/258) reported that they were not concerned, 23\% (60/258) were slightly unconcerned, 38\% (99/258) were neutral, 29\% (74/258) were slightly concerned, and 3\% (8/258) were extremely concerned. 
A Kruskal-Wallis test examined the effect of respondent technical proficiency on whether they are concerned about their privacy when using ESS tools and found significant differences ($H(X) = 120.1, p~<~$0.05).  
Participants' responses show that respondents with basic proficiency with technology were slightly concerned (43\% (18/42)), 
those with intermediate and advanced proficiency responded with neutral 38\% (62/162) and 37\% (20/54) of the time, respectively. 
 \end{blue}

%% file: 05-Interview.tex
\section{Semi-Structured Interview}
\label{sec:interview}
\new{Our online survey findings show that 68\% of respondents did not agree or strongly agree that they are concerned about privacy when using the ESS and 50\% are  satisfied with the ESS controls to manage their privacy. The semi-structured interview provides us with the opportunity to explore the low concern for privacy.} 
The semi-structured interview achieves three main objectives: (1) to verify participants' understanding relative to their reported ESS use and to gain deeper insights about the collected survey data; (2)~to gain insights into participants' behaviour when working with ESS on different devices (personal vs.\ corporate); and (3) to evaluate potential controls and indicators that could improve the privacy posture around ESS. All interview participants previously completed the first iteration of our online survey, and the interview was treated as an extension to the survey. 
\new{Note that since the interview participants were debriefed by the researchers, they were not asked to take the survey again.}

\subsection{Participants and Procedure}
Respondents that expressed interest in participating in the follow-up interview were shortlisted. This list was then categorized into five groups based on participants' 5-point Likert response to the perceived efficacy of the ESS privacy controls. We then chose participants from these groups with diverse education and knowledge of IT. 
We contacted 90 eligible participants \crs{from the first iteration of the online survey}, 34 responded,  and 22 participated in the interviews. 
During the interview, the researcher discussed all user-ESS interactions by the participant, but focused on one ESS that had the most diverse functionality  to elicit more meaningful responses, e.g., a SASE was preferred over a SSO solution.


Due to the COVID-19 pandemic, the interviews were conducted mostly 
online (using Microsoft Teams, Zoom, or Skype). If a 
participant chose to do the interview by phone, they were required to have Internet access during the interview to see our mockups. Their responses were documented by the researchers, which included feedback on the privacy controls and indicators developed in this work (see \S~\ref{subsubsec:interfaces}). Participants of the interview were paid \$20 and had a chance to win an Apple iPad. The interview questions were broadly categorized into the following groups, and required both categorical and free-form responses (see Appendix~\ref{app:interview}):

\begin{itemize}\itemsep0em
    \item \textbf{Software Installation and Configuration:} We requested participants to \new{recall} their initial experience with their ESS
    . We noted whether the software was pre-installed or pre-configured in addition to what information the ESS or the employer conveyed about data collection and their privacy.
    \item \textbf{Knowledge about Features and Data Collection:} We asked participants about the functionality and data collection practices of the software. If they were unaware of some functionality or data collection for the ESS, the interviewer bridged this gap while emphasizing that the functionality may not be enabled by the employer. 
    
    \item \textbf{Privacy Concerns:} We asked participants about their privacy concerns and noted how these differed between BYOD and corporate devices. 
    \item \textbf{Control over Privacy:} We asked participants whether they were able to voice their concerns, whether they were aware of the privacy controls and indicators that were provided by the ESS, and their thoughts on the efficacy of these privacy controls.
    \item \textbf{Privacy Controls and Indicators:} We consulted well-established principles for designing privacy notices~\cite{schaub2015design} to create mockups of security notices and indicators for \new{a generic} ESS. These were presented to the participants to get their preliminary feedback.
\end{itemize}


\begin{table}[t]
\caption{Demographics and inclusion criteria for interviews}
\label{tab:demographicsInterview}
\centering
\small\addtolength{\tabcolsep}{-2pt} \renewcommand{\arraystretch}{0.75}
\begin{tabular}{ccccccccc} \hline \addlinespace[1mm]
\multicolumn{9}{c}{{\bf n = 22}} \\ \hline \hline \addlinespace[1mm]
\multicolumn{9}{c}{{\bf Gender} } \\
\multicolumn{3}{c}{\emph{Man}}  & \multicolumn{3}{c}{\emph{Woman}}  & \multicolumn{3}{c}{\emph{Other}} \\
\multicolumn{3}{c}{16}    & \multicolumn{3}{c}{6}    &  \multicolumn{3}{c}{-}  \\ \hline  \addlinespace[1mm]
\multicolumn{9}{c}{{\bf Age (in years)}} \\
& \emph{18--25}               & \emph{26--30}               & \emph{31--35}               & \emph{36--40}               & \emph{41--45}               & \emph{46--50}               & \emph{50+} \\
& 6                 & 3                 & 3                 & 5                 & 1                 & 1              & 3 \\ \hline \addlinespace[1mm]
\multicolumn{9}{c}{{\bf Self reported proficiency in IT}} \\
\multicolumn{3}{c}{\emph{Basic}} & \multicolumn{3}{c}{\emph{Intermediate}} & \multicolumn{3}{c}{\emph{Advanced}} \\
\multicolumn{3}{c}{1} & \multicolumn{3}{c}{7} & \multicolumn{3}{c}{14}   \\  \hline  \addlinespace[1mm]
\multicolumn{9}{c}{{\bf Education or work in IT}} \\ 
\multicolumn{3}{c}{\emph{No}} & \multicolumn{3}{c}{\emph{Yes}} & \multicolumn{3}{c}{\emph{Prefer not to say}} \\
\multicolumn{3}{c}{6} & \multicolumn{3}{c}{16} & \multicolumn{3}{c}{-}   \\  \hline  \addlinespace[1mm]
\multicolumn{9}{c}{{\bf ESS provides privacy controls?} } \\
\multicolumn{3}{c}{\emph{(Strongly) Disagree}}  & \multicolumn{3}{c}{\emph{Neutral}}  & \multicolumn{3}{c}{\emph{(Strongly) Agree}} \\
\multicolumn{3}{c}{8}    & \multicolumn{3}{c}{8}    &  \multicolumn{3}{c}{6}  \\ \hline
\end{tabular}
\end{table}

\subsection{Results}
For the qualitative analysis of participants' responses during the interview, two researchers independently performed thematic analysis to identify themes.  \crs{Inter-rater agreement over identified themes was calculated and reported using Fleiss' Kappa. Then the identified themes were discussed and compared by researchers until consensus was reached.} 
Several other researchers in the field have used this approach (\crs{see McDonald et al.~\cite{Mcdonald2019internet} for details}). For the qualitative data from interviews, we provide representative quotes from participants for different themes. When presenting quotes, we also identify the number of participants who expressed that theme. 


Table~\ref{tab:demographicsInterview} summarizes the demographic information for participants of the semi-structured interviews. Among the participants,  six agreed or strongly agreed, eight were neutral, and eight disagreed or strongly disagreed that ESS provides sufficient privacy controls. Similarly, more participants reported advanced proficiency in IT. 
\new{We note imbalance for two aspects: more participants were male and more participants self-reported advanced proficiency in IT. Therefore, the IT proficiency limitations identified in \S~\ref{subsub:demographics} are also applicable for the interview results.}

\subsubsection{ESS Configuration and Policy Matters}
\label{subsec:interview:policy}
\subhead{Configuration} We asked participants about the type of device the ESS was installed on, and who installed/configured it. 15/22 of participants had it installed on corporate devices, 3/22 on personal devices, and 4/22 on both. Half of the participants (11/22) received their devices pre-installed with the software from the organization, while the rest set it up themselves. 

\subhead{ESS Privacy Policy} We asked participants if they recall seeing the privacy policy during installation or configuration.
5/22 of participants reported seeing some form of an agreement but did not read it, 6/22 reported having no knowledge of encountering a privacy policy (including participants with memory lapses), and 11/22 reported seeing and understanding some form of an agreement.
Participants who reported not reading the agreement had issues with how it was presented:
\begin{Q}
    \textit{``Obviously I'm not reading the full privacy policy, so [I] wish there was more communication on that [what is collected].''} (P15)
\end{Q}
Among the participants who reported not seeing a privacy policy, two attributed it to the remote nature of ESS.
\begin{Q}
    \textit{``They are all cloud-based, I am not sure if endpoints receive [information on what is collected]''} (P5)
\end{Q}
Finally, half of the participants who reported seeing and understanding some form of an agreement were unable to recall the details from the privacy policy: 
\begin{Q}
    \textit{``I think there were some disclosures, very general, can’t remember anything particular.''} (P6)
\end{Q}

 These results show that even with a higher-than-normal technical understanding among participants, there was uncertainty around how privacy notices were communicated.

\subhead{Company Policy and Training}  
We asked participants if they were provided any information by the organization as to what they can or cannot do with their corporate devices. Their responses show that 9/22 received instructions from their employer, while 12/22 received no information. 
The way this information was conveyed from the organization ranged from formal training \newer{(often through the employee handbook)}  to ``\textit{don't download stuff off the Internet}'' (P6).
Participants who reported not receiving this information, had the implicit understanding that they were required to ``\textit{keep things professional}'' (P15).

We asked participants what their employers told them about the functionality and need of the ESS.
\newer{Their responses were codified and an inter-rater agreement between the two researchers was substantial (Fleiss's $\kappa$ = 0.77).} Their responses show that 12/22 received no information, 5/22 received software specific information (e.g., VPN to provide secure access), 2/22 received organizational specific information (e.g., you need it to access organization's webservices through it), and finally, 3/22 received both organization and software specific information. 


\subsubsection{Privacy Perceptions}
\label{sec:interview_privacy}

\subhead{Privacy Concerns}
Before measuring privacy concerns for the ESS used by each participant, the interviewer identified the features of the ESS and the data requirements of these features from the ESS website and shared it with the participants. The interviewer also informed them that their employer may have chosen not to enable some of the advertised features. (The misconceptions uncovered during this disclosure are discussed in \S~\ref{subsec:misconceptions}.)
Participants were then asked about their privacy concerns. \new{Their responses were codified and an inter-rater agreement between the two researchers was almost perfect (Fleiss's $\kappa$ = 0.85)}. 11/22 of respondents reported being concerned about their privacy, 9/22 answered being not concerned, and 2/22 described being somewhat concerned. In addition to concerns around what was collected, three participants reported ambiguity about the nature of this data collection: 
\begin{Q}
    \textit{``I am concerned because I do not know exactly what is collected. Some [business] addresses that I query are related to my medical appointments. Is that history getting stored and who has access to that?''} (P22)
\end{Q}
Participants who were not concerned, understood that it was a company device and the data collection was to improve the security posture. Their responses were similar to:
\begin{Q}
    \textit{``Yes, I have an understanding that on the corporate device anything can be monitored or at least I use the device with this understanding.''} (P16)
\end{Q}

\subhead{Existing Controls and Indicators} 
To further understand the controls available to the participants to manage or learn about the data collection on the ESS, we asked participants about the privacy mechanisms provided by the software.
\newer{Their responses were codified and an inter-rater agreement between the two researchers was almost perfect (Fleiss's $\kappa$ = 0.90).} Their responses indicate that 10/22 of participants felt that they had no control available to them, 12/22 felt that they had some control. The interview data shows that these participants felt that the control was available through enabling or disabling the software. Despite the knowledge of 
this control, four participants were ambivalent to use it: 
\begin{Q}
    \textit{``There maybe a method to disable it but then the IT will not like it.''} (P2)
\end{Q}
One participant also commented on the reduced efficacy of this control due to their lack of knowledge on data collection:
\begin{Q}
    \textit{``I can enable or disable it. I would say that any control that is provided will be sensible if I know more about what is collected and stored.''} (P22)
\end{Q}

We also asked participants whether they found it challenging to remember if the ESS was active. 
\newer{Their responses were codified and an inter-rater agreement between the two researchers was almost perfect (Fleiss's $\kappa$ = 0.95).}
8/22 of them reported that it was something that they would not remember during their device usage, while the remaining 14/22 had no difficulty remembering it. 
Those who could not remember the ESS in the background cited their lack of a conscious thought about the ESS (\textit{``[I] am in the zone''} (P16) or \textit{``I have to go inside windows taskbar for it''} (P13)). Participants who did not have difficulty remembering it reported different strategies for their workflow: 6/14 reported not storing credentials of accounts unrelated to work on their corporate devices, and 2/14 used a second personal device in parallel.

\subsubsection{Effects of ESS on Device Usage Behaviour}
\label{sec:interview:usage_behaviour}
Next, we investigated how the knowledge of ESS data collection influences the device usage behaviour of participants. 10/22 of respondents reported a significant change in their behaviour, 3/22 reported a moderate change, and 9/22 reported no change. Their coping mechanisms are discussed in \S~\ref{sec:interview_privacy}, which include using their smartphone or another device to access the Internet, not saving credentials of accounts unrelated to work, and not copying personal files on the device with ESS. One participant also reported a coping mechanism that may not have been effective for the ESS:
\begin{Q}
    \textit{``[I] try not to do anything personal on the corporate device but sometimes it happens and when it does I make sure to log out of accounts and clear the history.''} (P12)
\end{Q}
For BYOD, if the ESS was used to access corporate resources, participants always reported disabling it after getting done with their tasks. However, they noted that during the access to corporate resources they have no other option but to keep it active. Two participants also reported disabling ESS on BYOD to protect the privacy of their family members.



\subsubsection{Perspective on Possible Solution}
We designed possible interfaces to communicate the privacy policy to the end-users and collected feedback. However, before seeking feedback, we asked participants what safeguards or solutions they desired. \newer{Their responses were codified and an inter-rater agreement between the two researchers was almost perfect (Fleiss's $\kappa$ = 0.81).} 
The most requested feature by 11/22 of participants was the ability to gain more knowledge about ESS privacy policy and better indicators showing when data collection was in progress. Three of these participants desired something similar to how smartphones conveyed and presented this information:
\begin{Q}
    \textit{``Communicate more like how smartphones do. I check that for apps before installing those [apps].''} (P18)
\end{Q}
6/22 of respondents desired the ability to enable/disable access to personal data of the users (\textit{``give control to [the] user for personal files and other personal data''} (P9)). 
3/22 of respondents expressed their desire for privacy policies/notices that are easier to comprehend along with the access to a summary of collected data.  

While participants provided some suggestions, often borrowing ideas from similar domains (e.g., permissions in smartphones), 5/22 did not provide any. While four of these five were satisfied with the state of affairs, one commented on the complex issue of providing privacy in enterprise security: 
\begin{Q}
    \textit{``[There are] inherent privacy violations [to providing security]. Without really understanding this [ESS functionality], it is difficult to approach and do anything about [privacy in ESS].''} (P11)
\end{Q}

We also explicitly asked participants if they would like access to the logs of data that are collected by the ESS. 
11/22 of participants said that they would like to have this access, 8/22 of participants did not, and 3/22 of participants were unsure. 
Half of the participants who wanted access, only wanted it once to understand what was collected:
\begin{Q}
    \textit{``I would like to see them once, to know what they are tracking. I don’t want this on a regular basis, after the first time, it would be useless, it won’t change anything.''} (P6)
\end{Q}
Two participants who reported not wanting the access did so because of their inability to react to it:
\begin{Q}
    \textit{``No, don't really care enough. Ignorance is bliss sometimes as I can't change what I use.''} (P17)
\end{Q}




\subsubsection{Privacy Notices and Indicators}
\label{subsubsec:interfaces}
Considerable research effort has focused on the design of privacy notices. We use the guidelines established by Schaub et al.~\cite{schaub2015design} for designing effective privacy notices. To this end, we design the following four different interfaces, which cover four timing aspects---at setup, just-in-time, periodic, and persistent. 
Note that we did not explore privacy controls or indicators for data collection sensors as different ESS may provide different controls depending on the ESS functionality, organization's policy, and sensors used for data collection. 

\subhead{(1) Privacy Notice} A privacy notice was designed to be displayed at the installation or configuration time (see Figure~\ref{fig:privacy_notice} in Appendix~\ref{app:interview}). It succinctly provided information on what is collected, with whom the collected data is shared, and how long the data is kept. We leverage findings of Gluck et al.~\cite{gluck2016short} on the representation and the framing of the privacy notice. 

\subhead{(2) Taskbar Indicator} To explore a persistent indicator, we created an indicator for the Windows Taskbar and macOS Menu Bar (see Figure~\ref{fig:taskbar} in Appendix~\ref{app:interview}). This indicator would always display the organization name to remind that it is a corporate device. Hovering the mouse over the indicator would give the user the option to view the ESS' privacy policy.  

\subhead{(3) Periodic Indicator} We explored a periodic indicator, where the user would be reminded of the ESS' presence by a toast shown repeatedly at a configurable frequency (see Figure~\ref{fig:toast} in Appendix~\ref{app:interview}). The toast would be displayed in the lower, center part of the screen for a few seconds, and then disappear. 

\subhead{(4) App Launch Indicator} Finally, we explored a just-in-time indicator, where the user would be reminded of the ESS' presence by a toast that will be shown at the launch of applications (see Figure~\ref{fig:toast} in Appendix~\ref{app:interview}). Other than the trigger logic, this toast is the same in appearance and functionality as the \emph{Periodic Indicator}. 

\vspace{4pt}
\noindent The participants were first shown all the interfaces and explained the functionality of each interface. They were then asked about the effectiveness of each interface, their likes, dislikes, feedback on potential improvements, and if they would use the interface if made available to them. 

\begin{figure}[t]
    \centering
    \includegraphics[trim={15mm 15mm 17mm 30mm},clip, width=0.95\linewidth] {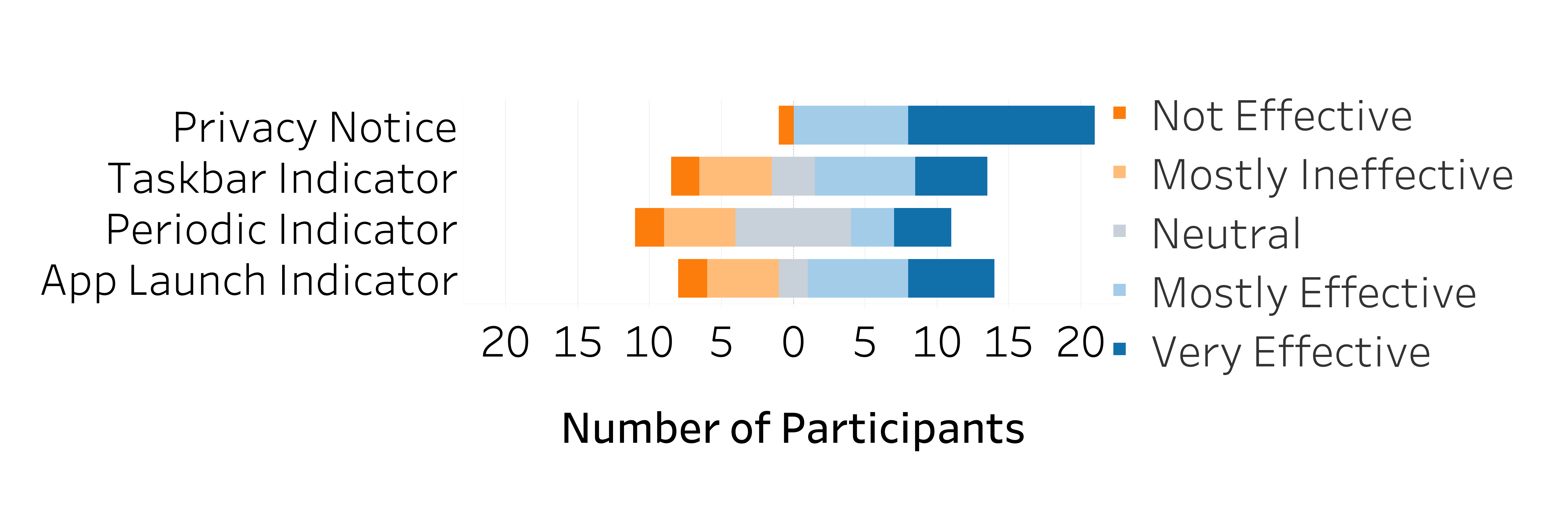}
    \caption{Mockups' efficacy rating by interview participants.}
    \label{fig:interface_efficacy}
\end{figure}

Figure~\ref{fig:interface_efficacy} shows the responses of the participants on a 5-point Likert scale. It shows that all but one participant felt that the \emph{Privacy Notice} was mostly or highly effective.
12/22 of them felt that the \emph{Taskbar Indicator} was mostly or highly effective.
Only 7/22 of participants felt that the \emph{Periodic Indicator} was mostly or highly effective while, 13/22 of them felt that the \emph{App Launch Indicator} was mostly or highly effective.

\newer{The participants' responses for likes, dislikes, and potential improvements were codified and all inter-rater agreements between the two researchers were substantial or higher (Fleiss's $\kappa$ = 0.70).} For the three indicators, most participants (12 or more) reported that they liked the improved visibility about the ESS operations. For \emph{App Launch Indicator}, three participants reported liking the ``appropriate timing'' of the toast. The main dislike reported for the \emph{Taskbar Indicator} was that it would ``eventually be passively ignored'' (5/22), whereas annoyance due to notifications was reported as major drawback for the \emph{Periodic Indicator} (11/22) and the \emph{App Launch Indicator} (4/22). Another point of consideration from 5/22 of the participants is that the \emph{App Launch Indicator} is not ideal for an application that may run for a longer period, resulting in the notification getting ``stale''. In terms of potential improvements, the majority of participants (11/22 or more) requested the ability to change location or frequency of the indicators. 4/22 of participants wanted a status indicator for the \emph{Taskbar Indicator} (i.e., data collection currently enabled/disabled).

In terms of adoption, 16/22 reported that they would like to use \emph{Privacy Notice}; 3/22 were unsure, and  3/22 did not want to use it. 10/22 of the participants wanted to use the \emph{Taskbar Indicator}; 5/22 were unsure, and 7/22 did not want to use it. Only 2/11 reported they would use the \emph{Periodic Indicator}, and 16/22 did not want to use it. For the \emph{New App Indicator}, 9/22 wanted to use it, while 7/22 were unsure. 
In \S~\ref{subsec:discussion_improvements}, we discuss how our preliminary explorations on these interfaces will help future research.




%% file: 06-Discussion.tex
\section{Discussion}
\label{sec:discussion}
Our online survey shows that 32\% of respondents are concerned about their privacy in presence of the ESS. When asked if they received information about data collection, only 20\% (116/585) of user-ESS interactions received ``clear'' information \new{from their employer and only 19\% (109/585) of user-ESS interactions received ``clear'' information from the ESS itself.} This low percentage brings into question the respondents' understanding of the ESS data collection practices \new{and how those practices could be made more transparent}. 
In terms of respondents' comfort, we note that the majority of participants were comfortable sharing different types of data and felt that the tools provided sufficient privacy controls. 
In contrast, the subset of participants that were invited for the interview expressed more concern about the ESS data collection. There are several possible reasons for this difference including:  (1) the subset of participants invited for the interviews was more balanced in terms of their satisfaction with the privacy controls of ESS; (2) during the interviews, the researcher debriefed the participants about the advertised features; and (3) during the interviews, the researcher focused on the ESS used by the participant that had more capabilities, thereby, a higher likelihood of data collection. 
In this section, we report on the sources of participants' misconceptions, and underlying issues related to trust and transparency on the part of the ESS and employers. We also provide suggestions for improving ESS privacy.

\subsection{Sources of Misconception}
\label{subsec:misconceptions}
Privacy perceptions of users 
are likely to be influenced by their understanding of what data is collected and who can see it. 
An incomplete or incorrect understanding is important to measure along with the sources of these misconceptions. To this end, before conducting the interview, we used the ESS' website to note the features provided, and data collected by the ESS used by each participant. During the interview, this information was requested again from the participant and we noted the gaps in their understanding. 

In terms of their understanding of ``what was collected'', we categorized their responses into three groups---correct understanding, underestimation, and overestimation of the collected data. (Note that we informed the participant about the possibility that the ESS may be listing a feature on their website but the participants' organization may not have it enabled.) 
Only 7/22 of participants had a correct understanding, 
2/22 of them overestimated, and 13/22 of them underestimated the functionality and associated data collection.
For instance, a popular ESS that provides endpoint application usage monitoring, DNS-layer protection, and traffic inspection service was reported only as a VPN software by all five participants who used this software. 
Similarly, one participant reported that the ESS \textit{``[is] scanning the [content of their] downloaded files''}(P14) when in fact, the ESS was monitoring the websites visited and checking incoming links and files prior to download.
This underestimation was a result of the lack of transparency and was coupled with the desire to understand why certain data should be collected: 
\begin{Q}
    \textit{``I am not sure why [ESS-name-redacted] needs to collect this information if I am not going on a website that is related to work.''} (P19)
\end{Q}
\begin{Q}
    \textit{``I still do not believe that it would be doing something beyond a VPN. I do not understand the need to do anything beyond it. But this software on my personal laptop raises my concern [level].''} (P20)
\end{Q}

It should be noted that the overestimation of the ESS functionality and data collection may result in trust issues. Two instances of note are provided below. For an enterprise email client on a BYOD, a participant misinterpreted Android's request for permission to ``access all files'' incorrectly as the organization requesting the ability to remotely wipe the device (the permission description said that this app may delete files that are not related to the app). 
One participant demonstrated a misconception regarding the extent of data collection when the ESS was turned off:
\begin{Q}
    \textit{``[I'm] not sure if you quit it [disable the ESS] it's still running in the background.''} (P15)
\end{Q}




We also investigated whether the information regarding software functionality and need provided by the employer (\S~\ref{subsec:interview:policy}) has any effect on the accuracy of the employees understanding. 
For the 13 participants who underestimated what was collected, 6/13 reported receiving no information from the employer while 7/13 reported receiving some information. 
For the 7 participants who had a correct understanding, 2/7 reported receiving no information while 5/7 reported receiving some information. For the two employees who overestimated, one reported receiving some information and the other did not. 
A Chi-squared test found no significant effect for the three understanding levels and whether the participants received the information from their employer  ($\chi^2(2) = 1.19, p~=~0.55$).

Another aspect to consider is the possibility that the organization's IT team may not know all the features that may be configured with the ESS, or different ways the collected data may be analyzed by the organization to which security services are outsourced.
\begin{Q}
    \textit{``... this tool is collecting information but I am not sure if my company knows that either. If they do, it should be their responsibility to inform me and other employees.''} (P19)
\end{Q}




\subsection{Trust of ESS and Employer}
\label{subsec:trust}
The lack of clear communication about the ESS functionality by the ESS or employer was often a source of mistrust. For instance, three participants self-discovered  that the ESS added certificates on their device without a warning and were alarmed since they were not informed about this feature: 
\begin{Q}
    \textit{``I had to pull some data from Linux console and it was failing. After much debugging I learned that my certificates had issue due to self-signed certificates causing the breaking of connections.''} (P15)
\end{Q}
\begin{Q}
    \textit{``My development IDE complained that its SSL certificate was being updated which was strange. Certificate that I had to accept was untrusted and I figured it [ESS' data collection] out that way.''} (P22)
\end{Q}
A notice to the users that certificates will be added and for what purpose would have avoided this issue. Several ambiguities that lead to the trust issues revolve around how the ESS worked, which were easily avoidable through communication:
\begin{Q}
    \textit{``I do not think this software should be installed on personal devices at least not without informing user as to what it does.''} (P2)
\end{Q}
\begin{Q}
    \textit{``[ESS-name-redacted] does analysis on [user] behaviour not exact file [scanning] so seems more secure [but] less personal.''} (P9)
\end{Q}

When asked about their privacy concerns, 5/22 employees reported that they trusted their employer and such data collection was necessary for security: 
\begin{Q}
    \textit{``No [privacy concerns]. [I] trust the company and they are doing nothing wrong on their corporate device.''} (P4)
\end{Q}

The trusting employees would also take necessary precautions to keep their sensitive personal data separate:

\begin{Q}
    \textit{``[I]  don't keep sensitive personal files on it, [and] trust company with the data stored on it.''} (P1)
\end{Q}

However, the response from 10/22 employees indicated a lack of trust, where the employers had the ability or knowledge but were withholding it from the employees:
\begin{Q}
    \textit{``Organization has to be truthful with the user but 99\% of the time they are not. [They] say they are collecting data but don’t explain the extent that they are.''} (P10)
\end{Q}

A comforting aspect is that even the employees that lacked trust in their employer or their ESS understood the need for such data collection, and identified open communication as a possible way to improve the trust:
\begin{Q}
    \textit{``I understand the company's need for security but it should not sound like my privacy for their security. They should at least inform what data is being collected.''} (P18)
\end{Q}



\subsection{Improving Privacy in ESS}
\label{subsec:discussion_improvements}
As discussed in \S~\ref{sec:intro}, data collection in the name of security has largely remained uncontested.
Our work provides the first insight into this ecosystem from the employees' perspective and  uncovers serious privacy concerns of employees. 
ESS products are installed on both corporate and BYOD devices. When users are required to use these devices with an incomplete understanding of the ESS data collection practices, the users' buy-in is less likely. While users may continue using the ESS, it would not be to improve the security posture of the organization but due to the lack of another option (\textit{``If I had a choice, I wouldn't want it [ESS], but I do want my job''} (P15)). The reduced buy-in may negatively affect the security posture of the organization~\cite{cranor2014better}.

As reported in \S~\ref{subsec:trust}, both trusting and untrusting employees understand the need of data collection by ESS, and do not contest the importance of security. However, most privacy concerns 
stem from less-than-optimal communication on the part of ESS developers and employers with respect to the ESS data collection (\S~\ref{subsec:misconceptions}). Clear communication through privacy notices is the first step towards enabling employees to have better understanding and control over their data.


Our \emph{Privacy Notice} mockup (\S~\ref{subsubsec:interfaces}) was deemed effective by most participants and helped them understand the data collection aspects better.
In contrast, the mockups that provided indicators received mix reviews. While the indicators that we experimented with improved the ESS and data collection visibility, our participants desired more. 
The mockups that we designed were the first attempt and the feedback provided by the participants (\S~\ref{subsubsec:interfaces}) should guide future research in this area.
ESS developers should provide users with more information on how to temporarily disable some functionality (e.g., traffic interception), and the employers should educate employees on the need of these features and the policy on whether the employees can disable them, and under what circumstances. 
Similarly, in the context of ESS, more research needs to be done to identify better indicators to inform users when data collection is in progress, insights into what is collected (potentially malicious vs.\ all behaviour), and other components of the data life cycle. 

Finally, one product packaging may not be suitable for both corporate and BYOD devices. As our findings show, the expectations for BYOD are different, and users desire more control over features and the data collected from their device. Respecting users' privacy without compromising security is essential and may help drive the adoption of the ESS:
\begin{Q}
    \textit{``I think there is definitely space for these tools to change the way they give users control. From a business perspective, they [ESS developers] want to sell their product. It would be great to have modular solutions and more consumer friendly.''} (P7)
\end{Q}

%% file: 07-Limitations.tex
\section{Limitations}
\label{sec:limitation}

\new{In \S~\ref{subsub:demographics}, we discuss limitations due to our convenience sampling and respondents performing the survey twice.} 
In addition, both phases of our study contain self-reported information, which may be influenced by the participants' memory, understanding, or subjective views. This may even have occurred in an effort to avoid embarrassment or to provide what they felt were favourable responses. 
\new{Our research methodology also could not adequately distinguish between the precise source of knowledge of each component of the ESS. For instance, sources of knowledge of different components could be different. Future research could overcome this with a longitudinal assessment of users' perspectives beginning with their initial interactions.} 


During our research, participants provided responses for privacy notices that were presented to them at the installation time, which may have been a long time ago (11/22 reported using their ESS for three or more years). As such, their recall of the details of the privacy notices may not have been accurate.
This limitation also highlights the need to ensure that easy access to the privacy policy is available even after the ESS is installed and configured. There was a month-long gap for some participants between their online survey responses and their semi-structured interview. This gap and the questions posed in the survey may have influenced the participants' privacy perceptions regarding ESS.

When measuring participants' understanding of the collected data and features of the ESS, we used the ESS website. However, employers may not have enabled all the advertised features. The lack of deployment of an advertised feature by the ESS may have skewed the measurements about participants' correct understanding of the functionality and the data collection, even following our disclosure during the interview. 


%% file: 08-Conclusion.tex
\section{Conclusion}
We conducted a survey (with 258 participants) and a semi-structured interview (with 22 participants) to understand employees' privacy perceptions and concerns when using ESS.
Our investigation provides evidence that there is a lot that is left to be desired by the end-user in terms of their privacy. This deficiency is due to the poorly designed notices and indicators from ESS developers, and the lack of communication from employers. Our study highlights how the two aforementioned issues result in misunderstanding and a lack of trust among several employees.
Finally, we suggest possible improvements in privacy notices and indicators that can be adopted by the ESS developers and communication of policy from the employers to mitigate the misconceptions of employees.
With the increasing adoption of ESS, our findings will help security and privacy researchers, ESS developers, and employers improve the privacy for employees.



%% file: 09-Appendix.tex
\appendix

\section{Appendix}

\subsection{Online Survey}
\label{app:online_survey}
\begin{enumerate} \itemsep0em\topsep=0em\partopsep=0em\parsep=0em
    \item \newer{If you previously participated in this study, have you since investigated Enterprise Security Tools more?}
    
    \item In which country do you currently reside?
    
    \item What is your age? \emph{(Age range dropdown)}

    \item What is your gender?
    
    \item \newer{Please select the highest level of education completed}
    
    \item Which of the following best describes your educational or occupational background?
    
    \item Which of the following best describes your level of proficiency with technology like smartphones or laptops?
    
    \item Please select security tools you have experienced. You may select up to 3 tools.
    \emph{Select from the list available in Appendix \ref{sfl:vendors} or provide their own}
    
    \item \newer{We want to test your attention during this survey. Therefore, to do so we please ask if you can select C as the answer for this question.}
    
\end{enumerate}
\newer{
\noindent
\emph{The next set of questions are repeated for each software selected:}
}
\newcommand{\SoftwarePlaceholder}{\emph{[software]} }

\begin{enumerate}\itemsep0em\topsep=0em\partopsep=0em\parsep=0em
  \setcounter{enumi}{9}
  \item Which of the following devices you use have \emph{[software]} installed? \\
  1) Personal Devices;
  2) Corporate Devices;
  3) Prefer not to say

  \item \newer{Select the industry in which you last used \emph{[software]}? \\
  1) Arts, entertainment, or recreation; 2) Education; 3) Financial; 4) Healthcare; 5)  Manufacturing; 6) Science/Research; 7)  Technology; 8) Other; 9) Prefer not to say}

  \item \newer{How many employees use \SoftwarePlaceholder at your organization?}
  
  \item\newer{ Do you currently use \emph{[software]}?\\}
    \textbf{[ Skip Q14 if Q13 is yes]}
  \item\newer{When was the last time you used \emph{[software]}?}
  
  \item \newer{How long have you been using \emph{[software]}?}
  
  \item \emph{[Software~|~Employer~|~Other Sources]} informed you about:
  \emph{Options for each: [Yes, clearly], [Yes, somewhat], [No], \newer{[Don't remember],} or [Prefer not to say]}
  \begin{compactitem}
      \item What it does (e.g., enables secure access to corporate email)
      \item Features (e.g., protection from viruses and network security)
      \item Data sources (e.g., files, network activity, and browser history)
  \end{compactitem}
  \newer{
  \textbf{[skip Q17 if Q16 response is No or Don't remember]}}
  \item For each of the following data sources, please specify if you were informed about \SoftwarePlaceholder usage of these resources by: \emph{[software]} itself, your employer, or self exploration:
  \emph{Options for each: [Software itself], [Employer], or [Self exploration],   \newer{[Don’t remember],} or [Prefer not to say]}\\
      Scanning personal files, 
      Uploading personal files, 
      Check the content of websites you visit, 
      Check the applications you use, 
      Tracking your location, 
      Other

  \item Please rate the following mechanisms in terms of your level of privacy comfort with enterprise security tools such as the tools you chose previously.
  \emph{\newer{Options for each (5-point Likert scale): [Not comfortable], [Slightly comfortable], [Comfortable], [Mostly comfortable], or [Very Comfortable]}}\\
      Scanning personal files, 
      Uploading personal files, 
      Check the content of websites you visit, 
      Check the applications you use, 
      Tracking your location, 
      Other
  
  \item Do you feel that the enterprise security tools such as the tools you chose previously provide you with sufficient controls to manage your privacy preferences?
  \newer{ \emph{Options for each (5-point Likert scale): [Strongly Disagree], [Disagree], [Neutral], [Agree], or [Strongly Agree]}}
  
  \item\newer{ Are you concerned about your privacy when using enterprise security tools such as the tools you chose previously?
  \emph{Option for (5-point Likert scale): [Extremely Concerned], [Slightly Concerned], [Neutral], [Slightly Unconcerned], [Not Concerned]}}
  
  \item \newer{Please rate your comfort as to who is able to see the data collected or accessed by enterprise security tools such as the tools you chose previously.
  \emph{Option for (5-point Likert scale): [Comfortable], [Slightly comfortable], [Neutral], [Slightly uncomfortable], [Not comfortable]}}
  \begin{compactitem}
    \item \newer{Computer Software}
    \item \newer{Organization's Analysts}
    \item \newer{Outside Organization's Analysts}
  \end{compactitem}
  
\end{enumerate}

\subsection{Semi-Structured Interview}
\label{app:interview}
\newcommand{\SurveySection}[1]{\noindent \underline{\emph{{#1}}}}

\begin{enumerate}\itemsep0em\topsep=0em\partopsep=0em\parsep=0em
	\item How was \SoftwarePlaceholder installed?
	\item Which device(s) do you use \SoftwarePlaceholder installed?\\
	    \emph{[Personal], [Corporate], or [Other]}
	\item Is \SoftwarePlaceholder required by your company, or a parent company/regulatory organization? 
\end{enumerate}

\begin{enumerate}\itemsep0em\topsep=0em\partopsep=0em\parsep=0em
	\setcounter{enumi}{3}
	\item Please describe the process you experienced in getting set up with \SoftwarePlaceholder (did you see any information related to privacy policy or what was collected)
	\item What do you remember was mentioned by your employer about \SoftwarePlaceholder in terms of functionality and its need?
	\item How long have you been using \emph{[software]}?
\end{enumerate}

\begin{enumerate}\itemsep0em\topsep=0em\partopsep=0em\parsep=0em
	\setcounter{enumi}{6}
	\item Please describe the functionality of \SoftwarePlaceholder 
	\item Please share examples of when you saw the functionality (including privacy/ data collection) of \SoftwarePlaceholder \emph{[Validate their understanding of the functionality and data collection of the said software]}
\end{enumerate}

\begin{enumerate}\itemsep0em\topsep=0em\partopsep=0em\parsep=0em
	\setcounter{enumi}{8}
	\item In your experience with \emph{[software]}, did you ever feel concerned about your privacy? Please explain.
	\begin{itemize}
		\item \emph{[Yes]} Was there an opportunity for you to voice your concerns? Did you feel it would make a difference? What if certain features could be disabled if sufficient employees were concerned about those (e.g., SSL connection breaking)? 
	\end{itemize}
	\item What implications come to mind with respect to \SoftwarePlaceholder and privacy?
	\item What privacy mechanism are you aware of in \SoftwarePlaceholder to provide control over your privacy? Overall, how much control do you feel over \emph{[software]}? 
		\emph{(5-point Likert scale)}
\end{enumerate}

\begin{enumerate}\itemsep0em\topsep=0em\partopsep=0em\parsep=0em
	\setcounter{enumi}{11}
	\item\emph{[For Interviewer] Before the interview, did the participant know about the privacy implications of \emph{[software]}?}
	\begin{itemize}
		\item{[Yes]: Do you change behaviour with respect to the \SoftwarePlaceholder on your computer in different settings?
		Have you noticed a change in patterns in your computing use when \SoftwarePlaceholder is active?
		How would the knowledge of the functionality of \SoftwarePlaceholder affect your behaviour and interactions with your devices?}
	\item{[No]: Would you change behaviour with respect to \SoftwarePlaceholder on your computer in different settings?}
	    \end{itemize}
	\item Do you disable \SoftwarePlaceholder in specific locations? 
	\item Do you find it hard to remember when it is activated?
\end{enumerate}

\begin{figure}[t]
    \centering
    \includegraphics[width=0.6\linewidth]{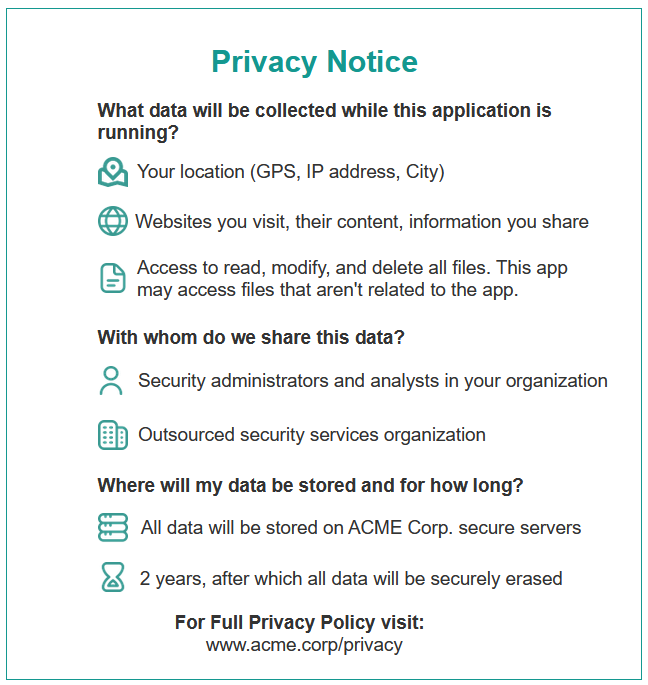}
    \caption{\emph{Privacy Notice} mockup, which is to be shown to the users at the installation or configuration time.}
    \label{fig:privacy_notice}
\end{figure}

\begin{figure}[t]
\centering
\begin{subfigure}{0.6\linewidth}{\includegraphics[clip,trim={30mm 0mm 0mm 25mm}, width=0.6\linewidth]{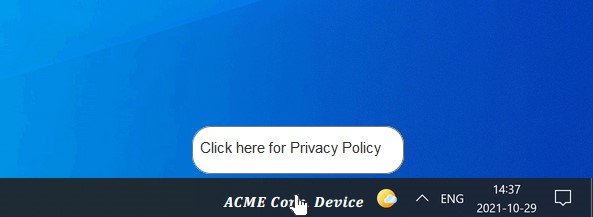}\caption{Windows Taskbar.}
\label{fig:windows_taskbar}} 
\end{subfigure}
\begin{subfigure}{0.6\linewidth}{\includegraphics[clip,trim={30mm 25mm 0mm 0mm}, width=0.6\linewidth]{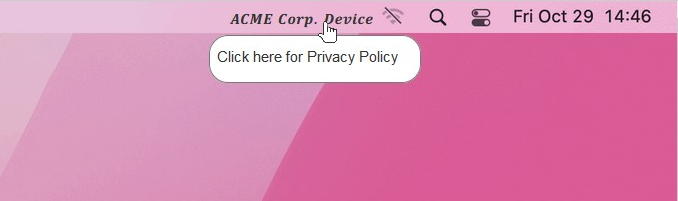}
\caption{macOS Menu Bar.}
\label{fig:macmenubar}}
\end{subfigure}
\begin{subfigure}{0.6\linewidth}{ \includegraphics[clip,trim={70mm 40mm 100mm 45mm},width=0.6\linewidth]{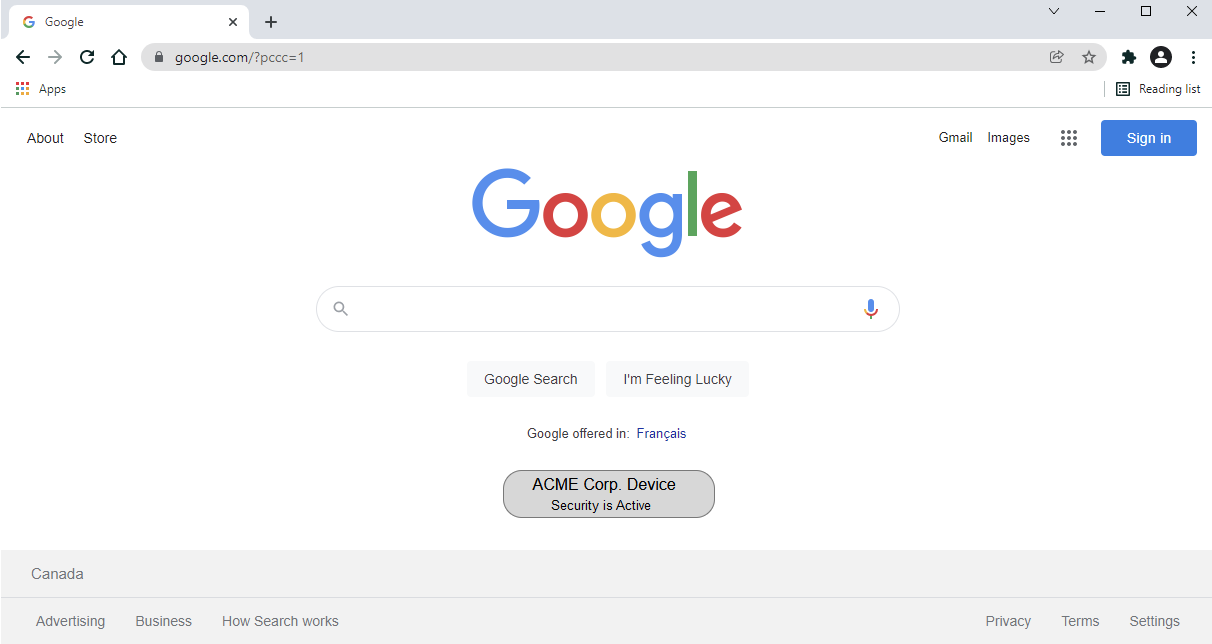}
\caption{Toast Notification.}
\label{fig:toast}}
\end{subfigure}
\caption{Mockups of persistent notification in the Taskbar or Menu Bar, and toast notification (note that these mockups are cropped).}
\label{fig:taskbar}
\end{figure}

\begin{enumerate}\itemsep0em\topsep=0em\partopsep=0em\parsep=0em
	\setcounter{enumi}{14}
	\item What end user controls would you like to have present to feel more comfortable while using the device with \emph{[software]}?
	\item Please rate the following mechanisms in terms of your level of privacy comfort:
			\emph{(5-point Likert scale)}\\
			Scanning personal files, 
			Uploading personal files, 
			Check the content of websites you visit, 
			Check the applications you use, 
			Tracking your location
	\item Would you like access to the logs of what was collected from you?
	\item Are there any other safeguards or solutions that you know of, or you can think of, that would better the end user experience and trust with the software?
	\item How would your responses differ if it was your personal device vs. corporate provided device?
		\emph{[No difference for either device], [More protection for personal], [Less protection for organizational, Other]}

\end{enumerate}
\textbf{[Privacy Notice (Figure \ref{fig:privacy_notice}), Taskbar (Figures \ref{fig:windows_taskbar} and \ref{fig:macmenubar}), Periodic (Figure \ref{fig:toast}), New App Launch (Figure \ref{fig:toast})]}
\begin{enumerate}\itemsep0em\topsep=0em\partopsep=0em\parsep=0em
	\setcounter{enumi}{19}
	\item Please rate the effectiveness of the interfaces in communicating the privacy notice.
	\emph{(5-point Likert scale)}\\
			Privacy Notice, 
			Taskbar (Figures, 
			Periodic, 
			New App Launch
	\item Please rank the interfaces in terms of your preference:
	Taskbar, Periodic, New App Launch
	\end{enumerate}
\begin{enumerate}\itemsep0em\topsep=0em\partopsep=0em\parsep=0em
\setcounter{enumi}{21}
    \item What do you Like, Dislike, or would change about the following:
	Taskbar,
	Periodic,
    New App Launch
    
	\item Which of the mock-up interfaces would you use if they were available on your corporate devices. How would this differ on your personal devices?
		   
\end{enumerate}

\subsection{ESS List}
\label{sfl:vendors}

Akamai Enterprise Application Access, Appgate, Barracuda CloudGen Access, BlackBerry Optics, Cato SASE, Cisco SASE, Cisco Secure Endpoint, Cortex XDR by Palo Alto Networks, Cybereason Defense Platform, Falcon by Crowdstrike, FireEye Endpoint Security (HX), Fortinet SASE (FortiSASE), Kaspersky Endpoint Detection and Response (KEDR), Malwarebytes Endpoint Detection and Response, McAfee Endpoint Threat Defense and Response, Microsoft Defender for Endpoint (MDE), Microsoft Intune Company Portal, Netkope, Okta, OneLogin, PaloAlto Prisma Access, Panda Adaptive Defense 360 by WatchGuard, Perimeter 81 SASE, Singularity Platform by SentinelOne, Sophos Intercept X Advanced with EDR, Symantec Advanced Threat Protection, Trend Micro XDR, Twingate, VMWare Carbon Black, VMWare SASE, Zscaler SASE.

